\documentclass[aps,prmaterials,showkeys,twocolumn,nofootinbib,superscriptaddress]{revtex4-1}
\usepackage{natbib}
\usepackage{amssymb}
\usepackage{color}
\usepackage{amsfonts}
\usepackage{textcomp}
\newcommand{\ped}[1]{\ensuremath{_{\rm #1}}}
\newcommand{\apex}[1]{\ensuremath{^{\rm #1}}}
\definecolor{blue}{rgb}{0,0,0}
\definecolor{magenta}{rgb}{0,0,0}
\usepackage{float}
\usepackage[caption = false]{subfig}
\usepackage{graphicx}

\begin{document}

\title{Ambipolar suppression of superconductivity by ionic gating in optimally-doped BaFe\ped{2}(As,P)\ped{2} ultrathin films}

\author{Erik Piatti}
\affiliation{Department of Applied Science and Technology, Politecnico di Torino, 10129 Torino, Italy}
\author{Takafumi Hatano}
\affiliation{Department of Materials Physics, Nagoya University, Furo-cho, Chikusa-ku, Nagoya 464-8603, Japan}
\author{Dario Daghero}
\affiliation{Department of Applied Science and Technology, Politecnico di Torino, 10129 Torino, Italy}
\author{Francesco Galanti}
\affiliation{Department of Applied Science and Technology, Politecnico di Torino, 10129 Torino, Italy}
\author{Claudio Gerbaldi}
\affiliation{Department of Applied Science and Technology, Politecnico di Torino, 10129 Torino, Italy}
\author{Salvatore Guastella}
\affiliation{Department of Applied Science and Technology, Politecnico di Torino, 10129 Torino, Italy}
\author{Chiara Portesi}
\affiliation{INRiM -- Istituto Nazionale di Ricerca Metrologica, 10135 Torino, Italy}
\author{Ibuki Nakamura}
\affiliation{Department of Materials Physics, Nagoya University, Furo-cho, Chikusa-ku, Nagoya 464-8603, Japan}
\author{Ryosuke Fujimoto}
\affiliation{Department of Materials Physics, Nagoya University, Furo-cho, Chikusa-ku, Nagoya 464-8603, Japan}
\author{Kazumasa Iida}
\affiliation{Department of Materials Physics, Nagoya University, Furo-cho, Chikusa-ku, Nagoya 464-8603, Japan}
\author{Hiroshi Ikuta}
\affiliation{Department of Materials Physics, Nagoya University, Furo-cho, Chikusa-ku, Nagoya 464-8603, Japan}
\author{Renato S. Gonnelli}
\email{renato.gonnelli@polito.it}
\affiliation{Department of Applied Science and Technology, Politecnico di Torino, 10129 Torino, Italy}

\begin{abstract}
Superconductivity (SC) in the Ba-122 family of iron-based compounds can be controlled by aliovalent or isovalent substitutions, applied external pressure, and strain, the combined effects of which are sometimes studied within the same sample. Most often, the result is limited to a shift of the SC dome to different doping values. In a few cases, the maximum SC transition at optimal doping can also be enhanced. In this work, we study the combination of {\color{blue}charge doping} together with isovalent P substitution and strain, by performing ionic gating experiments on BaFe\ped{2}(As$_{0.8}$P$_{0.2}$)\ped{2} ultrathin films. We show that the polarization of the ionic gate induces modulations to the normal-state transport properties that can be mainly ascribed to {\color{blue}surface charge doping}. We demonstrate that {\color{blue}ionic gating} can only shift the system away from the optimal conditions, as the SC transition temperature is suppressed both by electron and hole doping. We also observe a broadening of the resistive transition, which suggests that the SC order parameter is modulated non-homogeneously across the film thickness, in contrast with earlier reports on {\color{blue}charge-doped} standard BCS superconductors and cuprates.

\end{abstract}


\maketitle

\section{Introduction}\label{sec:introduction}
Barium-122 (BaFe\ped{2}As\ped{2}) is the parent compound of one of the most widely studied classes of Fe-based superconductors, thanks to the availability of high-quality single crystals and thin films. Substitutional doping suppresses the spin-density-wave phase typical of the parent compound, and promotes the emergence of a superconducting dome \cite{RotterPRL2008, KatasePRB2012, SefatPRL2008, KasaharaPRB2010, DhakaPRL2011, Reticcioli2017}. Different chemical elements can be used as dopants by partially substituting either Ba, Fe, or As atoms. In the first case, the Ba reservoir can be doped by alkali-metal (indirect hole doping \cite{RotterPRL2008}) or rare-earth substitution (indirect electron doping \cite{KatasePRB2012}). In the second case, the FeAs layers are directly doped by, e.g., substituting Fe with Co (direct electron doping \cite{SefatPRL2008}) or Ru (isovalent doping \cite{DhakaPRL2011, Reticcioli2017}). In the third case, As atoms are substituted with P atoms (isovalent doping \cite{KasaharaPRB2010}). These isovalent substitutions strain the crystal structure of the parent compound, leaving the charge density unaffected (chemical pressure doping \cite{KasaharaPRB2010, DhakaPRL2011, Reticcioli2017}). All of these methods lead to the onset of superconductivity (SC).

However, while the isovalency of P and As atoms (or Ru and Fe atoms) guarantees that the carrier density of BaFe\ped{2}(As,P)\ped{2} is left unchanged, alkali-metal and Co substitutions lead to simultaneous charge doping and strain on the crystal structure, making it impossible to completely disentangle their effects on the SC state. In this framework, {\color{blue}the surface charge doping induced by ionic gating constitutes an ideal tool to investigate the problem, since it allows tuning the surface carrier density of a material while reducing distortions to the crystal structure with respect to standard charge doping via chemical substitution \cite{UenoRevie2014}. 
Ionic gating exploits the ultra-high} electric field at the interface between a solid and an electrolyte to accumulate surface charge densities in excess of $10^{15}$ cm\apex{-2} in the so-called electric double layer (EDL) \cite{DagheroPRL2012, LiNature2016}. Such large densities allow tuning the electric transport properties even in highly conductive systems, such as thin films of metals \cite{DagheroPRL2012, TortelloApsusc2013} and BCS superconductors \cite{ChoiAPL2014, PiattiJSNM2016, PiattiPRB2017}, or thin flakes of transition-metal dichalcogenides \cite{LiNature2016, XiPRL2016}. Moreover, ionic gating has been proven to be a very effective tool to explore the phase diagram of Fe-based superconductors, controlling the magnetic/Mott phase transition in TlFe\ped{1.6}Se\ped{2} \cite{KatasePNAS2014} and the SC transition in FeSe\ped{0.5}Te\ped{0.5} \cite{ZhuPRB2017}, as well as triggering the development of a high-temperature SC phase in FeSe \cite{ShiogaiNatPhys2016, LeiPRL2016, HanzawaPNAS2016, MiyakawaPRM2018} and FeSe\ped{0.8}Te\ped{0.2} \cite{KounoArXiv2018}. {\color{blue}Gate-induced lithiation has also been reported to very effectively tune the phase diagram of FeSe \cite{LeiPRB2017} and (Li,Fe)OHFeSe \cite{LeiPRB2016}.}

In this work we concentrate on ultrathin ($\sim10$ nm) films of optimally P-doped BaFe\ped{2}As\ped{2} {\color{blue}to allow for an efficient gate-tuning of their physical properties despite the strong electrostatic screening typical of metallic systems. We employ the ionic gating technique to induce surface charge doping levels} up to $\sim3.5\times10^{14}$ cm\apex{-2}, aiming to explore the effect of a doping method ``orthogonal'' to the isovalent chemical one. Our films show a suppression of the critical temperature $T_c$ for both positively and negatively induced charge densities. This suggests that the films optimized for the highest $T_c$ with respect to the isovalent P content and the strain induced by the substrate, are also intrinsically optimized with respect to the charge doping. This unexpected result may help to better understand this intriguing class of superconductors and act as a guide for further fundamental studies. 

\bigskip

\section{Device fabrication}

BaFe\ped{2}(As$_{1-x}$P$_{x}$)\ped{2} films were grown by molecular beam epitaxy (MBE) on MgO substrates at $850^{\circ}$C under ultra high vacuum (base pressure $\sim 10^{-9}$ mbar) according to the procedure described in Ref.\onlinecite{KawaguchiSUST2014}. Vapors of Ba, Fe, and As were supplied from pure metal sources in Knudsen cells. P vapors were supplied from a GaP cell, which was equipped with a Ga trap to obtain an almost pure P flux. The P vapor pressure was adjusted in order to control the P content $x$, while a stable growth rate $\simeq1.67$ nm min\apex{-1} was obtained by controlling the As, Fe and Ba fluxes \cite{KawaguchiSUST2014}. The growth time for the film batch from which the field-effect devices were fabricated was set to $6$ minutes in order to obtain a film thickness $\simeq10$ nm as per the calibrated growth rate. The resulting composition of the thin films was investigated by electron probe micro-analysis (EPMA), confirming the optimal P content $x=0.21$ and ensuring that no Ga was incorporated during the growth. Note that, in BaFe\ped{2}(As,P)\ped{2} films epitaxially grown on MgO, the optimal doping value is shifted to lower P content values \cite{KawaguchiSUST2014} with respect to single-crystals \cite{KasaharaPRB2010}. This is because epitaxial films grown on MgO substrates develop an in-plane tensile strain that shifts the SC dome to lower P contents with respect to single crystals \cite{KawaguchiSUST2014}.

\begin{figure*}
\begin{center}
\includegraphics[keepaspectratio, width=0.7\textwidth]{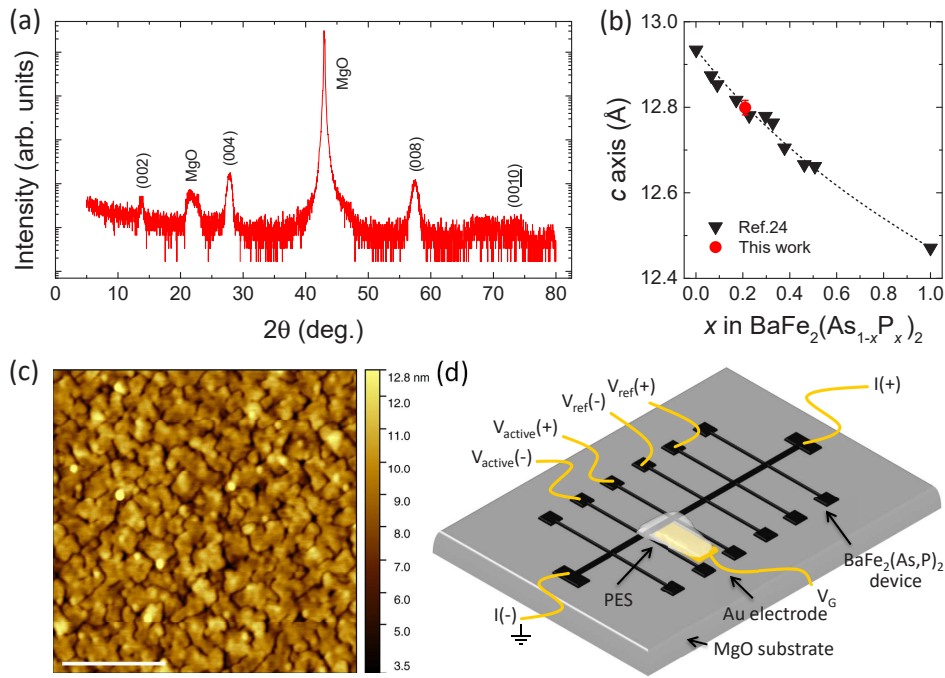}
\end{center}
\caption {
(a) XRD spectrum of a $\simeq$ 10 nm thick BaFe\ped{2}(As\ped{0.79}P\ped{0.21})\ped{2} film.
(b) Out-of-plane lattice constant $c$, determined from XRD, as a function of the P content $x$. Red circle is obtained from the spectrum in (a). Black down triangles are adapted from Ref.\onlinecite{KawaguchiSUST2014}. Black dashed line is a guide to the eye.
(c) AFM topographic image acquired in tapping mode of the surface of the same thin film. Root mean square height is $S_q\simeq1.5$ nm, much smaller than the nominal film thickness. Scale bar is $1\,\mathrm{\mu m}$.
(d) Sketch of a BaFe\ped{2}(As,P)\ped{2} device with the electrical connections required for double-channel four-wire resistance and gating measurements.
} \label{figure:fabrication}
\end{figure*}

Subsequently, the structural properties were probed both via X-ray diffraction (XRD) and atomic force microscopy (AFM). Fig.\ref{figure:fabrication}a shows a representative XRD pattern of our $10$ nm-thick, optimally doped films. XRD measurements were performed by means of a Cu K$\alpha$ X-ray source, and indicate that the thin film grew on the MgO substrate with a strong orientation along the $c$ axis even in the presence of a significant lattice mismatch. The out-of-plane lattice constant $c$ was obtained through the $(002)-(00\underline{10})$ reflections in the out-of-plane $\theta-2\theta$ spectrum. When plotted against the P content determined from EPMA (see Fig.\ref{figure:fabrication}b), the $c$ lattice parameter shows excellent agreement with thicker films ($\sim 100$ nm) grown via the same method \cite{KawaguchiSUST2014}. Fig.\ref{figure:fabrication}c shows a representative $3\times3\,\mathrm{\mu m}^2$ AFM topography scan acquired with a Bruker Innova\textsuperscript{\textregistered} scanning-probe microscope in tapping mode. The MBE growth resulted in a granular film, with well-defined grains having a mean equivalent square size $\simeq0.1\,\mathrm{\mu m}$ and featuring sharp edges between each other. This is in contrast with thicker ($\sim 50$ nm) films grown via the same method, where the grains coalesce in continuous, overlapping terraces with a much larger mean equivalent square size $\simeq0.4\,\mathrm{\mu m}$ \cite{DagheroApsusc2017}. The surface roughness, estimated via the root mean square height $S_q\simeq 1.5$ nm, is much smaller than the nominal film thickness.

After being characterized, thin films were patterned in Hall-bar shape (see Fig.\ref{figure:fabrication}d) by photolithography and ion milling (Ar gas, $10^{-3}$ mbar, extraction voltage $400$ V, anode current density $1.2$ mA cm\apex{-2}); each pair of adjacent voltage contacts defines a channel that is $1$ mm long and $0.3$ mm wide. The liquid precursor to the polymer electrolyte system (PES) was then drop-casted on one of the channels in the controlled atmosphere of a dry room and UV-cured. The resulting geometry is sketched in Fig.\ref{figure:fabrication}d, and is chosen to allow for the simultaneous measurement of two different channels on the same device: the active channel, covered by the electrolyte and acting as the working electrode of the electrochemical cell, and an ungated reference channel. A gold leaf dipped into the PES acts as the gate counter electrode.

Our PES of choice consists of a soft, cross-linked polymer matrix with a glass phase transition below $240\,\mathrm{K}$ and containing a solvated salt; the salt ions are not bound to any specific molecule, increasing the resulting EDL capacitance with respect to standard polymer electrolytes \cite{DagheroPRL2012}. The polymeric matrix is composed by a mixture of BEMA dimethacrylate oligomer, i.e. bisphenol A ethoxylatedimethacrylate (average $M_{w}\sim1700$ daltons, Sigma Aldrich) and PEGMA mono methacrylate based reactive diluent, i.e. poly(ethylene glycol) monomethyl ether monomethacrylate (average $M_{w}\sim500$ daltons, Sigma Aldrich) in 7:3 ratio along with 3 wt \% of free radical photo initiator (Darocur 1173, Ciba Specialty
Chemicals). 10 wt\% of lithium bis(oxalato)borate salt (LiBOB) was then added as the active source of ions.

Our choice to use LiBOB over more standard salts, such as lithium bis(trifluoromethanesulfonyl)imide (LiTFSI) or lithium perchlorate (LiClO\ped{4}) was dictated by {\color{blue}its superior chemical and electrochemical stability \cite{TaubertJTES2010, XuESSL2005, JowJES2004}:} {\color{magenta}compared to standard conductive salts, LiBOB shows several advantages, including a higher thermal stability and less corrosive hydrolytic decomposition products \cite{TaubertJTES2010}. Most importantly, however, polarization of LiBOB-based electrolytes is known to lead to the formation of a stable solid electrolyte interface (SEI) film at the electrode surface, preventing interactions with potentially aggressive species dissolved in the electrolyte \cite{TaubertJTES2010}. 
Indeed, attempts at using other active salts to gate BaFe$_2$(As,P)$_2$ films resulted either in permanent electrochemical modification of the film surface, or outright etching and dissolution of the film into the electrolyte, leading to device failure.}  
On the other hand, LiBOB-based electrolyte solutions generally exhibit significantly suppressed ionic conductivities with respect to more standard active salts \cite{TaubertJTES2010}, as well as poorer performances below room temperature \cite{XuESSL2005}. {\color{blue}Further details about the stability and performance of the LiBOB-based PES can be found in the Supplemental Material \cite{Supplemental}.} 

\begin{figure}
\begin{center}
\includegraphics[keepaspectratio, width=\columnwidth]{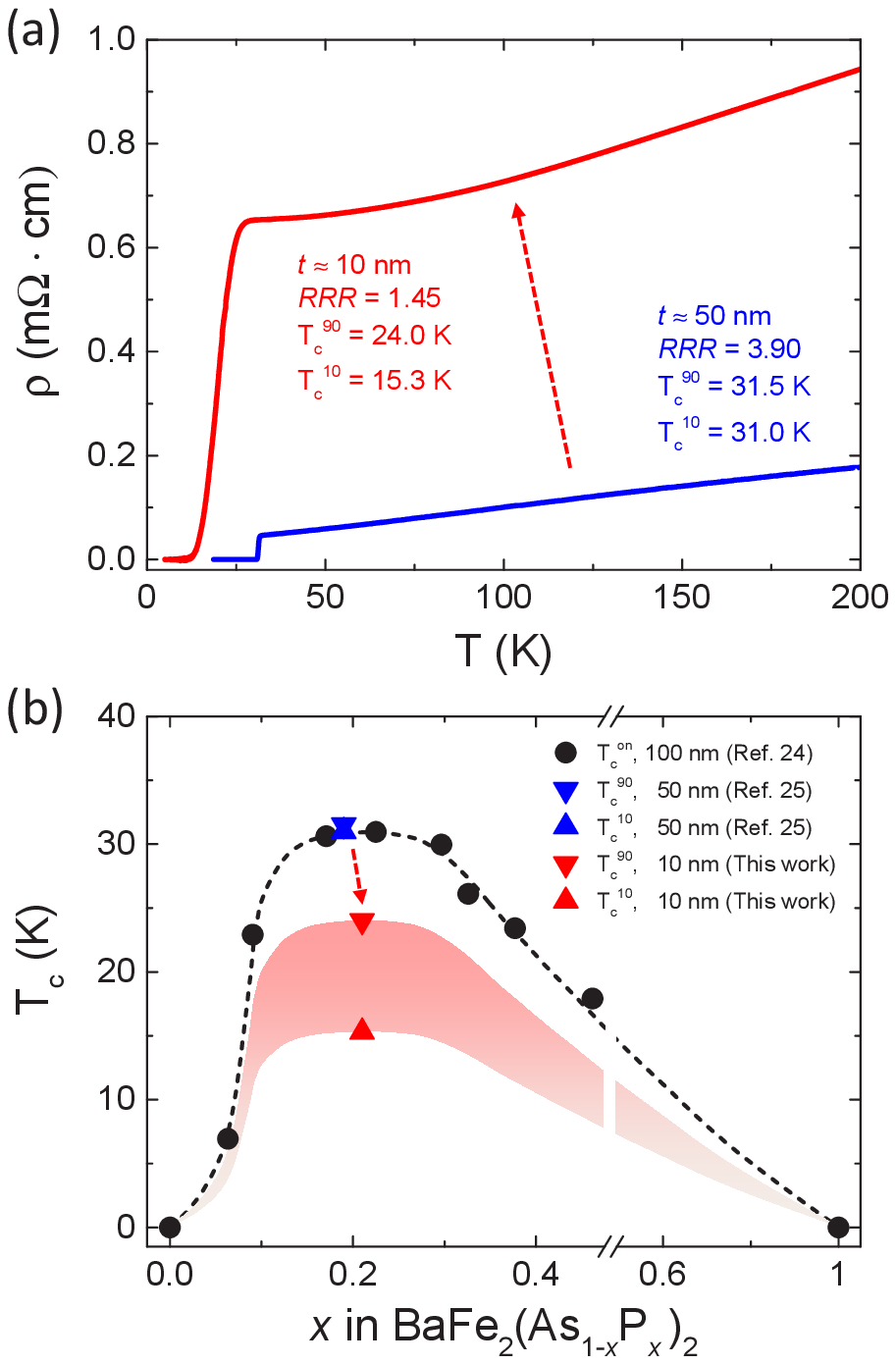}
\end{center}
\caption {
(a) Resistivity, $\rho$ vs. temperature, $T$, for two optimally-doped BaFe\ped{2}(As,P)\ped{2} films of different thickness. The solid red line refers to the $10$ nm thick devices fabricated in this work, the solid blue line to a $50$-nm film of comparable P content ($x=0.19$) and identical growth process from Ref.\onlinecite{DagheroApsusc2017}. SC transition temperatures and residual resistivity ratios $RRR = \rho(200\,\mathrm{K}) / \rho(32\,\mathrm{K})$ are also reported.
(b) SC transition temperatures, $T_c$, vs. P content $x$. Red and blue down (up) triangles refer to $T_c^{90}$ ($T_c^{10}$) for the $10$ and $50$ nm thick films respectively, obtained from the curves shown in panel (a). Black circles are $T_c^{onset}$, adapted from Ref.\onlinecite{KawaguchiSUST2014}. The dashed black line is a guide to the eye. The shaded band sketches the possible broadening of the SC transition in the thinner films away from optimal doping.
} \label{figure:ungated_transport}
\end{figure}

\section{Results}

Complete devices were rapidly transferred in the chamber of a Cryomech\textsuperscript{\textregistered} pulse-tube cryocooler and left to degas for at least 1 hour in high vacuum ($\lesssim 10^{-5}$ mbar) to remove any trace of water absorbed by the PES. Four-wire resistance ($R$) measurements were performed by applying a small DC current of a few $\mu$A to the current contacts of the device with a low-noise Keithley 6221 current source, and measuring the longitudinal voltage drops V\ped{active}, V\ped{ref} across the active and reference channels with a Keithley 2182 nanovoltmeter. Common-mode offsets (such as thermoelectric voltages) were removed by source-drain current inversion. A preliminary $R$ vs. $T$ characterization was performed on each device before PES drop-casting by cooling the sample down to 5 K and letting it slowly heat up to room temperature via the small residual thermal leak to the environment; the $R(T)$ curves reported in the following were all measured during the slow, quasi-static heat-up of the sample.

Fig.\ref{figure:ungated_transport}a shows the $T$ dependence of the resistivity, $\rho(T)$, of one of our $10$ nm thick devices, in the absence of the ionic gate (solid red line). For comparison, we also show the $\rho(T)$ data of a $50$ nm thick epitaxial film of similar P content ($x=0.19$), grown with the same method (solid blue line, adapted from Ref.\onlinecite{DagheroApsusc2017}). With respect to the thicker film, our device shows strongly enhanced values of $\rho$ in the entire $T$ range. This mainly stems from its larger saturating $\rho$ at low $T$ - as marked by its smaller residual resistivity ratio $RRR = \rho(200\,\mathrm{K}) / \rho(32\,\mathrm{K})$. Moreover, the SC transition temperature (Fig.\ref{figure:ungated_transport}b) is significantly suppressed in the thinner sample ($T_{c}^{90}$ is reduced by $7.5$ K, $T_{c}^{10}$ by almost $15$ K), while the width of the SC transition is enhanced ($T_{c}^{90}-T_{c}^{10}$ increases from $0.5$ K to $8.7$ K). Here, $T_c^{10}$ and $T_c^{90}$ indicate the $T$ values at which $\rho$ reaches $10$ and $90$\% of its value in the normal state $\rho(32\,\mathrm{K})$. On the other hand, the $T_c$ of the thicker sample agrees very well with those of $100$ nm thick epitaxial films of similar composition \cite{KawaguchiSUST2014}. This indicates that the thickness reduction from $50$ to $10$ nm is responsible for the suppression and broadening of the resistive transition at optimal doping, which is very likely to occur across the SC dome as a function of P content (as sketched in the red shaded band in Fig.\ref{figure:ungated_transport}b). 

This marked suppression and broadening of the resistive transition with decreasing thickness could be either an intrinsic feature of Ba-122 thin films, as in the case of YBa$_2$Cu$_3$O$_{7-\delta}$ \cite{GoodrichPRB1997, TangSUST2000}, or instead be due to the specific growth conditions of our samples. The first interpretation can be supported by the observation of a suppressed and broadened transition in Ba(Fe,Co)$_2$As$_2$/STO superlattices when the thickness of the Ba-122 layers approaches $\sim 12$ nm \cite{LeeNatMater2013}, as well as the absence of SC in Ba(Fe,Co)$_2$As$_2$ films less than $3$ nm thick \cite{EomPrivCom}. The second interpretation can instead be associated with the different surface morphology between our $10$ nm thick film and the $50$ nm thick film of Ref.\onlinecite{DagheroApsusc2017} as evidenced by AFM: indeed, granular growth of Ba-122 thin films can strongly suppress and broaden the resistive transition, as well as strongly increase the low-$T$ resistivity, especially in presence of in-plane misalignment between the grains \cite{LeeNatMater2010}. Moreover, the presence of sharp boundaries between the grains in the thinner film is likely to locally suppress the SC order parameter, leading to poor superfluid connectivity and weak-link SC behavior \cite{LikharevRMP1979, ClaassenAPL1980}. The thicker films, on the other hand, would be more robust against these issues since their thickness is large enough for the grains to coalesce in quasi-continuous terraces \cite{DagheroApsusc2017}. {\color{blue}Additionally, the transport properties of thicker films would be less sensitive to the partial surface oxidation which is unavoidable when BaFe\ped{2}As\ped{2} films are removed from high vacuum \cite{PlecenikAPL2013}.} While we cannot rule out a contribution from the first mechanism, we deem that this second interpretation is more likely to account for the behavior of our films.

Even if ultrathin films display a broadened SC transition, minimizing sample thickness is necessary to effectively tune the physical properties of any metallic system via the electric field effect. This is due to the very efficient electrostatic screening associated to their large carrier density, which confines any perturbation to few atomic layers from the surface even in the presence of the large electric fields typical of the ionic gating technique \cite{PiattiApsusc2018NbN}. Minimizing the sample thickness is even more necessary in the case of superconducting films, where any field-induced modification of $T_c$ becomes suppressed with increasing film thickness due to the proximity to the unperturbed bulk \cite{PiattiPRB2017, UmmarinoPRB2017}.

After this preliminary characterization, we modulated the charge doping in two different devices, which were fabricated from $10$ nm thick BaFe\ped{2}(As\ped{0.8}P\ped{0.2})\ped{2} films grown in the same batch to ensure full consistency between the measurements. Charge doping was induced by applying, at $T=290$ K, a gate voltage $V_G$ between the negative current contact and the gate counter electrode. This temperature was chosen to minimize the chances of electrochemical interactions, while avoiding an excessive reduction in the ionic conductivity of the LiBOB salt due to its well-known poor performance at lower temperatures \cite{XuESSL2005}. Both the application of $V_G$ and the measurement of the gate current $I_G$ flowing through the electrolyte were performed with a Keithley 2410 source measure unit (SMU).

\begin{figure}
\begin{center}
\includegraphics[keepaspectratio, width=\columnwidth]{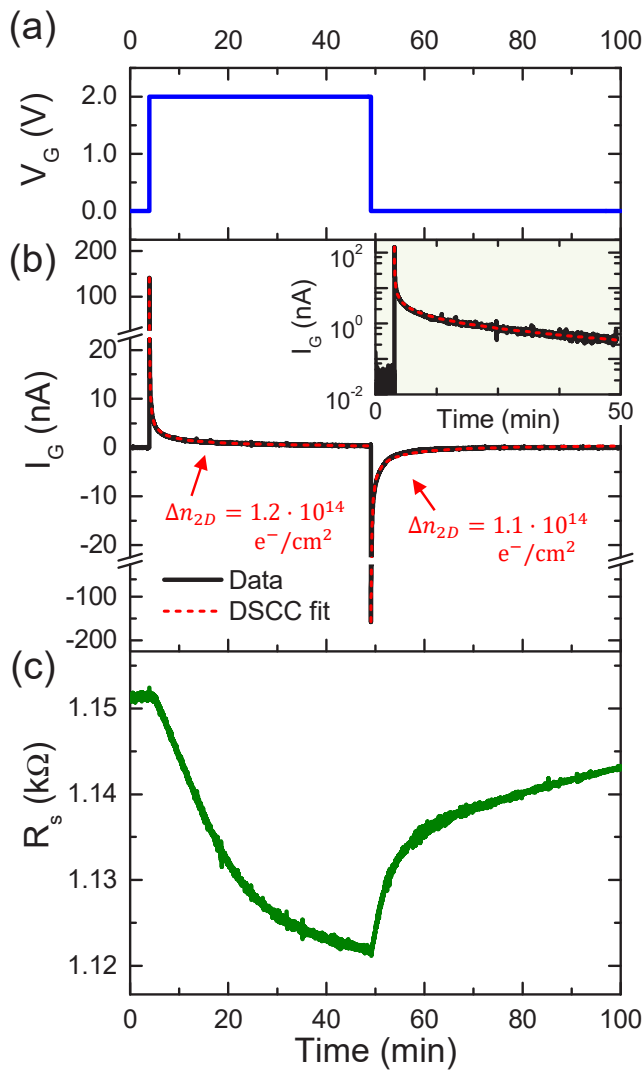}
\end{center}
\caption {
Gate voltage $V_G$ (a), gate current $I_G$ (b) and active channel sheet resistance $R_s$ (c) vs. time for typical step-like application and removal of $V_G$ at $T=290$ K. Panel (b) includes also the fits to the gate current (dashed red lines) according to the DSCC models, and the resulting estimations of the induced charge density per unit surface $\Delta n_{2D}$. Inset to panel (b) shows the same data of the first $50$ min of (b) in semilogarithmic scale.
} \label{figure:gating}
\end{figure}

Fig.\ref{figure:gating} shows a typical response of the active channel to a step-like application and removal of a chosen value of $V_G$. We always applied, and removed, $V_G$ in a step-like fashion (Fig.\ref{figure:gating}a) to allow for double-step chronocoulometry (DSCC): this is a well-established electrochemical technique that allows a reliable determination of the amount of charge induced at the surface of the active channel due to the build-up of the EDL \cite{Scholtz2010, DagheroPRL2012, PiattiJSNM2016, PiattiPRB2017}. Fig.\ref{figure:gating}b shows the recorded $I_G$ flowing through the electrolyte (solid black line). The dashed red lines represent instead the fit to the experimental data in the charge and discharge processes, obtained within the DSCC model. The total amount of accumulated charge, $\Delta n_{2D}$, as determined by DSCC for both processes is also indicated. In the following, $\Delta n_{2D}>0\,(<0)$ will refer to electron accumulation (electron depletion). Fig.\ref{figure:gating}c shows the response of the sheet resistance $R_s$, which, in the case of electron doping, consists in a reduction. It is immediately apparent that the electrolyte is characterized by very long transient times ({\color{magenta}on the order of} tens of minutes), both for the charge and the discharge of the EDL capacitor{\color{blue}, as can be observed also in the very slow vanishing of $I_G$ when plotted in semilogarithmic scale (see inset to Fig.\ref{figure:gating}b). The sizeable values of $\Delta n_{2D} \sim 10^{14}\,\mathrm{cm^{-2}}$, together with the very long transient times, might suggest that the observed charge doping requires mechanisms beyond the electrostatic polarization of the electric double layer. However, an estimation of the Debye length $\lambda_D$ for our PES composition \cite{Supplemental, ChazalvielBook, RusselBook, SengwaPI2000} gives $\lambda_D \approx 0.1\,\mathrm{nm}$, indicating that the EDL can be described by the compact layer approximation and its capacitance $\sim 10\,\mathrm{\mu F\,cm^{-2}}$ \cite{Supplemental, BlackAMI2017, ChazalvielBook, RusselBook, SengwaPI2000}, is large enough to account for the measured values of $\Delta n_{2D}$. Furthermore, the ionic conductivity of our LiBOB-based PES, as determined by electrochemical impedance spectroscopy (EIS) \cite{PorcarelliJPS2017}, is small ($\lesssim 10^{-7}\,\mathrm{\Omega^{-1}\,cm^{-1}}$) \cite{Supplemental}. This in turn results in an expected gate relaxation time $\tau_G \sim 10$ min \cite{Supplemental} according to the model presented in Ref.\onlinecite{ZhouJAP2012}, and comparable to the transients exhibited by our devices. Therefore, we can conclude that the dominant contribution to charge doping in our devices is likely electrostatic, while contributions from electrochemical effects are, if present, below our detection limit. Additional details concerning the gate charging dynamics, as well as the results of linear sweep/cyclic voltammetry, EIS and X-ray photoelectron spectroscopy (XPS) experiments, can be found in the Supplemental Material \cite{Supplemental}.}

Further insight into the interplay between charge doping and gate-induced modulation of the electric transport properties can be obtained from the scaling behavior of the variation of $R_s$ with the induced charge density $\Delta n_{2D}$. For single-band metallic films of thickness $t$, and assuming that the effective mass and scattering lifetime of the charge carriers remain unperturbed, a trivial free-electron calculation gives \cite{DagheroPRL2012,TortelloApsusc2013}:
\begin{equation}
\frac{\Delta R}{R'} = \frac{R(\Delta n_{2D})-R(0)}{R(\Delta n_{2D})} = -\frac{\Delta n_{2D}}{n_{3D,0}\cdot t}
\label{eq:deltaR_single_band_constant_tau}
\end{equation}
where $n_{3D,0}$ is the unperturbed carrier density per unit volume. That is, when the only effect of the application of $V_G$ is the accumulation/depletion of charge carriers, $\Delta R/R'$ should be scaling linearly with the induced charge density $\Delta n_{2D}$, with a sign that depends on whether the unperturbed charge carriers are electrons or holes.

\begin{figure}
\begin{center}
\includegraphics[keepaspectratio, width=\columnwidth]{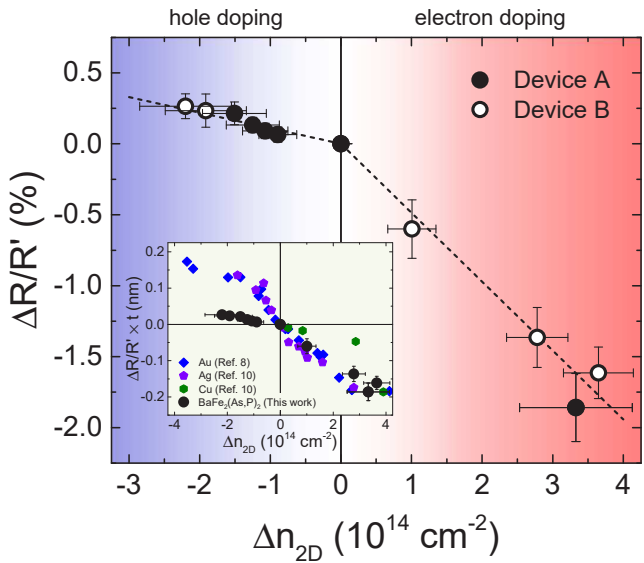}
\end{center}
\caption {
Normalized resistance variation, $\Delta R/R'$, vs. induced charge density, $\Delta n_{2D}$, at $T=290$ K. Hollow and filled circles refer to measurements obtained on two different samples from the same growth batch. Dashed lines are linear fits to the experimental data.
Inset: dependence of $\Delta R/R' \times t$, where $t$ is the film thickness, on $\Delta n_{2D}$, for different metallic materials. Blue diamonds refer to Au thin films (adapted from Ref.\onlinecite{DagheroPRL2012}), violet pentagons to Ag thin films and green hexagons to Cu thin films (adapted from Ref.\onlinecite{TortelloApsusc2013}). Black circles are the same data shown in the main figure.
} \label{figure:gated_scaling}
\end{figure}

Fig.\ref{figure:gated_scaling} shows that, for both electron and hole doping, the experimentally measured $\Delta R/R'$ is indeed a linear function of $\Delta n_{2D}$ and thus consistent with a gating behavior dominated by charge doping. Here, vertical and horizontal error bars are determined by comparing the values of $\Delta R$ and $\Delta n_{2D}$ between the application and removal of a given $V_G$ value, as showcased in Fig.\ref{figure:gating}. Since BeFe$_2$(As,P)$_2$ is a multiband system, we expect contributions to $\Delta R$ coming from electronic and holonic bands to have opposite sign and, thus, to partially cancel each other out. Since we observe the overall slope of the linear behavior to be finite and negative, we conclude that the conductivity of the system is dominated by quasiparticles carrying a negative charge. As we show in the inset to Fig.\ref{figure:gated_scaling}, both the sign and the magnitude of the modulations are comparable with previous results obtained via electrostatic gating on other metallic thin films \cite{DagheroPRL2012, PiattiJSNM2016, TortelloApsusc2013, ChoiAPL2014, PiattiPRB2017} and are thus consistent with a modulation of the density of charge carriers in the system.

On the other hand, this linear scaling of $\Delta R/R'$ on $\Delta n_{2D}$ exhibits a clear asymmetry between electron and hole doping, with the former being significantly more effective in tuning the conductivity in the system. This asymmetry between electron and hole doping was not observed in ion-gated metallic thin films \cite{DagheroPRL2012, TortelloApsusc2013}. It has however been reported when very surface-sensitive materials are ion-gated, such as black phosphorus \cite{SaitoACSNano2015}, and single-layer \cite{MinSciRep2017} and few-layer \cite{PiattiApsusc2016, Gonnelli2dMater2017} graphene, where it has been found consistent with a starkly different efficiency between cations and anions in introducing extra scattering centers during the build-up of the EDL. Indeed, by dropping the assumption of a constant quasiparticle scattering lifetime, the same free-electron calculation of Eq.\ref{eq:deltaR_single_band_constant_tau} gives:
\begin{equation}
\frac{\Delta R}{R'} = -\frac{\Delta n_{2D}}{n_{3D,0}\cdot t} \cdot \frac{\tau(\Delta n_{2D})}{\tau(0)}
\label{eq:deltaR_single_band_variable_tau}
\end{equation}
where $\tau(\Delta n_{2D})$ and $\tau(0)$ are the doping-dependent and unperturbed quasiparticle scattering lifetimes respectively. According to this interpretation, the scaling shown in Fig.\ref{figure:gated_scaling} indicates that $\tau|_{\Delta n_{2D}<0} < \tau|_{\Delta n_{2D}>0}$, i.e. the BOB\apex{-} anions (or the SEI formed during the electrolyte polarization with $V_G<0$) are more effective than the Li\apex{+} cations in introducing extra scattering centers at the surface of our devices.

{\color{blue}We now discuss some possible sources of doping beyond pure electrostatic polarization of the EDL. A first source is specific adsorption, where ions in the electrolyte move beyond their solvation shell and come into direct contact with the electrode surface \cite{ChazalvielBook}; we observed possible hints to this behavior in cyclic voltammetry tests at large positive $V_G$ \cite{Supplemental}. A second source is ion intercalation in the bulk of the film, which could be promoted by the layered structure of BaFe\ped{2}As\ped{2}. Intercalation by the BOB\apex{-} anion can be easily ruled out, since its large size would lead to device failure due to delamination of the layered structure \cite{YuNatNano2015}. The Li\apex{+} cation would not encounter this issue: however, bulk intercalation in EDL transistors is usually activated above certain threshold values of the gate electric field and associated with a sudden increase in disorder \cite{YuNatNano2015, PiattiAPL2017, PiattiApsusc2018MoS2}, and we would therefore expect it to lead to large deviations from the linear scaling of $\Delta R/R'$ with $\Delta n_{2D}$ that we instead observe in Fig.\ref{figure:gated_scaling}. Additionally, while intercalation can be readily obtained in materials of the 11 family of Fe-based compounds \cite{LeiPRB2016, LeiPRB2017}, it is strongly hindered in the 122 family by the presence of the positively charged, alkaline-earth charge reservoirs: namely, the Sr-122 parent compound is known to be prone to intercalation while the Ba-122 one is not \cite{HosonoMaterToday2018}, possibly due to the smaller spacing between the layers in the latter. In general, alkali metals (in particular K) can penetrate in the Ba-122 structure only when they \textit{substitute} Ba atoms, leading to the well-known SC dome induced by indirect hole doping \cite{RotterPRL2008}. XPS analysis we carried out in large-area films (see Supplemental Material for details \cite{Supplemental}) also does not reveal significant Li\apex{+} incorporation in the lattice, whether at the surface or in the bulk, and is consistent with the literature \cite{PlecenikAPL2013, deJongPRB2009, McLeodJPCM2012}. Overall, we deem the chance of significant Li\apex{+} intercalation in the lattice to be unlikely. 

A third source of charge doping beyond pure electrostatic polarization could arise from reversible distortions of the crystal lattice such as field-induced displacements of the Ba\apex{2+} charge reservoirs from their equilibrium positions, a mechanism similar to the one recently proposed to account for the long relaxation times of ion-gated ZrNCl \cite{ZhangCPL2018}. Permanent deintercalation of the Ba\apex{2+} charge reservoirs can be ruled out owing to the insensitivity of the Ba XPS spectrum to the gating process \cite{Supplemental}. Finally, a fourth source could arise from field-assisted protonation of the lattice, a mechanism which has been reported in the cases of SrCoO\ped{3} \cite{LuNature2017} and SrTiO\ped{3} \cite{LiArXiv2018} due to electrolysis of residual water traces in the gate electrolyte; while our voltammetry tests \cite{Supplemental} do not reveal peaks clearly attributable to water hydrolysis, a quantitative investigation of this contribution to charge doping requires \textit{in operando} characterization of the film XRD pattern and goes beyond the scope of the present paper.

Overall, we can safely conclude that these contributions to charge doping are likely secondary with respect to the electrostatic polarization of the EDL, as evidenced by the excellent linear scaling of $\Delta R/R'$ with $\Delta n_{2D}$ in the transport experiments, combined with the dedicated linear-sweep/cyclic voltammetry and XPS characterizations we discuss in the Supplemental Material \cite{Supplemental}.}

We now focus on how ionic gating can tune the SC transition of BaFe\ped{2}(As,P)\ped{2} thin films. We thus consider several $R$ vs. $T$ curves for different values of $\Delta n_{2D}$, for $T\leq30$ K, and determine their corresponding $T_{c}^{10}$, $T_{c}^{50}$, and $T_{c}^{90}$: this has the added advantage of allowing us to quantify any {\color{magenta}gate-induced} broadening of the SC transition. For each threshold, we then define the $T_c$ shift measured during the i-th thermal cycle as the difference between the $T_c$ of the active and reference channels:
\begin{equation}
\delta T_{c}(\Delta n_{2D})|_i=T_{c,act}(\Delta n_{2D})|_i-T_{c,ref}|_i .\label{eq:deltaTc}
\end{equation}

In general, $\delta T_{c}(\Delta n_{2D}=0)\neq0$ due to sample inhomogeneity, and $T_{c,ref}|_i\neq T_{c,ref}|_j$ due to slight differences in the heating rate between different measurements, or imperfect thermal contact between the sample and the thermometer. Using a ``differential measurement'' of $T_c$ on two channels of the same device sidesteps both issues \cite{PiattiJSNM2016, PiattiPRB2017}. The doping-dependent $T_c$ shift is then defined as:
\begin{equation}
\Delta T_{c}(\Delta n_{2D})=\delta T_{c}(\Delta n_{2D})-\delta T_{c}(0) .\label{eq:DeltaTc}
\end{equation}

\begin{figure}
\begin{center}
\includegraphics[keepaspectratio, width=\columnwidth]{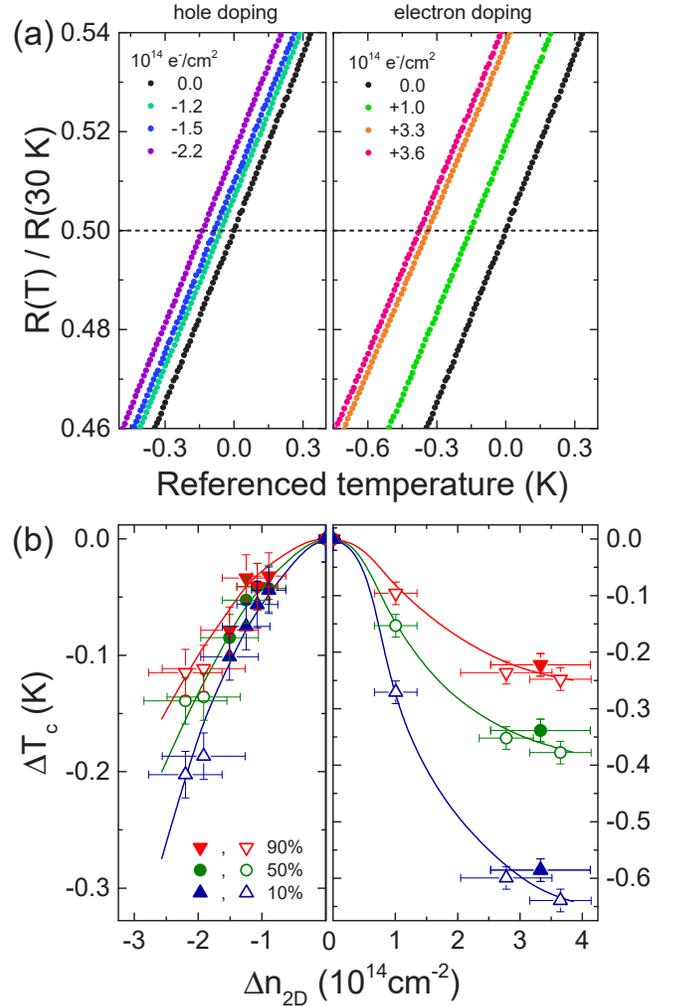}
\end{center}
\caption {
(a) Normalized resistance $R(T) / R(30\,\mathrm{K})$ of the active channel, vs. referenced temperature $[T-T_c^{ref}]_{\Delta n_{2D}} - [T_c^{active}-T_c^{ref}]_{0}$, in the vicinity of $T_{c}^{50}$. Left panel shows electron depletion (hole doping), right panel electron accumulation (electron doping). (b) $T_c$ shift, $\Delta T_{c}$, vs. induced charge density, $\Delta n_{2D}$, calculated for $T_c^{90}$ (red down triangles), $T_c^{50}$ (green circles), and $T_c^{10}$ (blue up triangles). Hollow and filled symbols refer to measurements obtained on two different samples from the same growth batch.
} \label{figure:gated_transport}
\end{figure}

Fig.\ref{figure:gated_transport}a shows the effect of different charge doping values on the $R$ vs. $T$ curve, close to the midpoint of the SC transition. On the vertical scale, $R(T)$ is normalized by its value at $30$ K, i.e. $R (T) / R (30\,\mathrm{K})$. On the horizontal scale, $T$ is referenced to $T_c^{50}$ in the reference channel, i.e. $[T-T_c^{ref}]_{\Delta n_{2D}} - [T_c^{active}-T_c^{ref}]_{0}$ \cite{PiattiPRB2017}.

Both electron and hole doping result in $\Delta T_{c}<0$, with electron doping leading to stronger $T_c$ suppression at comparable doping levels. Furthermore, most of these $T_c$ shifts are reversible, i.e. they disappear upon heating up the devices to $290\,\mathrm{K}$, setting $V_G = 0\,\mathrm{V}$ and waiting for a suitably long time. Due to the slow ion dynamics associated to the LiBOB-based PES, this could require several {\color{magenta}tens of minutes}. In the very few cases where complete reversibility was not observed, the original $\delta T_{c}(\Delta n_{2D}=0)$ could anyway be recovered by removing the PES and rinsing the device in ethanol. These results are again consistent with the tuning of $T_c$ mainly occurring via electrostatic charge doping \cite{DagheroPRL2012, PiattiPRB2017}. If more complex electrochemical interactions do give some contributions, these do not lead to a permanent modification of the SC properties of the films{\color{blue}, as evidenced also by the XPS analysis of the gated film surface \cite{Supplemental}}: as such, we tentatively ascribe these occurrences of incomplete reversible behavior {\color{blue}either} to long-term trapping of charged ions in the SEI formed by LiBOB decomposition at the sample surface \cite{TaubertJTES2010}{\color{blue}, or to a metastable distortion of the crystal lattice induced by the gate electric field \cite{ZhangCPL2018}}.

In Fig.\ref{figure:gated_transport}b we summarize all the reversible $\Delta T_{c}$ values as a function of $\Delta n_{2D}$ for the two devices, using the $T_c^{90}$, $T_c^{50}$ and $T_c^{10}$ criteria. Any variation of the {\color{blue}charge doping} with respect to the pristine value results in a reduction of $T_c$. Interestingly, the foot of the SC transition, $T_c^{10}$, is much more sensitive to doping than the onset, $T_c^{90}$, resulting in a broadening of the resistive transition for increasing charge doping. This behavior is starkly different from earlier observations in thin films of the standard electron-phonon superconductor NbN \cite{PiattiJSNM2016, PiattiPRB2017} and the high-$T_c$ superconductor YBa\ped{2}Cu\ped{3}O$_{7-\delta}$ \cite{FeteAPL2016}, where the SC transition was \emph{rigidly shifted} as a function of {\color{blue}surface} charge doping. Such a rigid shift is the fingerprint of a homogeneous modification of the SC state in the thin film \cite{PiattiPRB2017}{\color{blue}, while spatially-dependent modulations to the SC order parameter can lead to significant broadening \cite{PaolucciPRApplied}. Therefore,} the observation of a doping-induced broadening of the resistive transition suggests that the gate-induced {\color{blue}surface charge doping} gives rise to an inhomogeneous perturbation of the SC order parameter across the film thickness, consistent with the very small out-of-plane coherence length of BaFe\ped{2}(As,P)\ped{2} \mbox{($\xi_c(0)\simeq 11\div15$ \AA) \cite{RamosPRB2015}}.

{\color{blue}We also note that, for comparable doping levels, the broadening is significantly more pronounced for electron rather than hole doping, suggesting a different length scale over which the SC order parameter is suppressed in the two cases. Owing to the small value of $\xi_c(0)$, this perturbation likely follows the charge doping profile across the film thickness, suggesting a different spatial dependence of hole and electron doping in the out-of-plane direction, and independently of whether the charge doping is induced by electrostatic gating or electrochemical modification of the surface. In the former case, the asymmetry in the broadening could be ascribed to a different electrostatic screening length between electron and hole induction, similar to the case of gated MoS\ped{2} \cite{BrummePRB2015}.} 

In principle, a further contribution to the broadening of the resistive transition may also arise from a gate-induced increase in disorder. This may occur due to the introduction of extra scattering centers via the accumulation of ions at the film surface, an issue which is well-documented across a wide range of different materials \cite{GallagherNatCommun2015, OvchinnikovNatCommun2016, PiattiApsusc2016, Gonnelli2dMater2017, SaitoACSNano2015, MinSciRep2017, PetachACSNano2017, LuPNAS2018, PiattiNanoLett2018}. However, in this case we expect this contribution to be negligible, since the width of the resistive transition of BaFe\ped{2}(As,P)\ped{2} is known to be very robust against the introduction of extrinsic disorder, both in the case of single-crystals \cite{MizukamiNatCommun2014} and epitaxial thin films \cite{DagheroApsusc2017, DagheroSUST2018}. {\color{blue}More importantly, the scaling of $\Delta R/R'$ with $\Delta n_{2D}$ in the normal state indicates that disorder is more efficiently introduced by hole doping with respect to electron doping: therefore, if the broadening of the resistive transition was dominated by gate-induced disorder, one should observe a larger broadening upon hole doping and a smaller broadening upon electron doping. Since the opposite behavior is observed instead, we conclude that the contribution to the broadening caused by gate-induced disorder, if present, is minor with respect to the one introduced by the spatially-dependent modulation of the SC order parameter along the $c$ axis.}

\section{Discussion}

Earlier works combined substitutional doping and applied pressure in order to control the SC state in the Ba-122 family. In many of these works, the type and amount of chemical substitution were fixed for each sample, while the external physical pressure was employed as a quasi-continuous knob to control the SC properties \textit{in-situ}. Applying an external pressure to Co- \cite{ColombierSUST2010}, K- \cite{TorikachviliPRB2008, HassingerPRB2012} and P-doped \cite{KlintbergJPSJ2010} bulk samples resulted in a $T_c$ enhancement only in the underdoped regime, while optimally and overdoped samples featured a $T_c$ suppression. That is, the external pressure ``shifted'' the SC dome to lower doping values. The very same behavior was also observed when direct electron doping and chemical pressure were combined in co-doped Ba(Fe$_{1-y}$Co$_y$)$_2$(As$_{1-x}$P$_x$)$_2$ samples \cite{ZinthEPL2012}. A pressure-driven enhancement of $T_c$ across almost the entire phase diagram was only reported for the aliovalent substitution of Ba with La, which provides indirect electron doping to the FeAs layers \cite{KatasePRB2013}.

In this context, the charge doping induced by ionic gating {\color{blue}could potentially} be considered somewhat akin to indirect doping via substitution of the charge-reservoir atoms -- i.e., the dopants do not directly substitute the Fe atoms in the FeAs layers. Hence, at the optimal chemical pressure achieved via isovalent P substitution, one may {\color{blue}hypothetically} expect (i) $T_c$ to be suppressed by hole doping ($\Delta n_{2D}<0$), similarly to the aliovalent K substitution; (ii) $T_c$ to be enhanced by electron doping ($\Delta n_{2D}>0$), similarly to the aliovalent La substitution. On the other hand, it is important to note that physical and chemical pressure -- while they do have a very similar effect on the SC properties of Ba-122 \cite{KlintbergJPSJ2010} -- are not completely equivalent. Namely, P substitution is known to introduce a sizable uniaxial component \cite{KlintbergJPSJ2010} and results in a starkly different dependence of the Fe-As bond length on P content with respect to applied physical pressure \cite{ZinthEPL2012}. As such, it is not obvious that the interplay between indirect electron doping and pressure would be the same for physical and chemical pressure.

Indeed, our results show that ion-gate-induced electron doping, P-induced chemical pressure and the substrate-induced strain interact in a qualitatively similar way to direct electron doping and applied pressure (physical or chemical). That is, the chosen P content $\simeq 0.2$ (combined with the substrate-induced tensile strain) already optimizes SC, and any further change to the carrier density brings the system away from these optimized conditions. Further experiments on underdoped BaFe\ped{2}(As,P)\ped{2} films will be required to assess whether ionic gating is able to enhance SC in the underdoped regime, or if the suppression of SC extends to the entire phase diagram.

Furthermore, our results make clear how -- unlike in the cases of FeSe \cite{ShiogaiNatPhys2016, LeiPRL2016, HanzawaPNAS2016} and FeSe$_{1-x}$Te$_x$ \cite{ZhuPRB2017, KounoArXiv2018} -- the charge doping provided by ionic gating has a negligible impact on the SC state with respect to the different types of substitutional doping \cite{RotterPRL2008, KatasePRB2012, SefatPRL2008, KasaharaPRB2010} and applied pressure \cite{IshikawaPRB2009}. This finding confirms that, in the Ba-122 family, SC is much more strongly tied to modifications of the crystal structure than to the carrier density in the system \cite{ZinthEPL2012, KimberNatMater2009}. We note, however, that the real effectiveness of charge doping in modulating the $T_c$ of our samples may actually be underestimated in these experiments. This is because{\color{blue}, in the absence of bulk electrochemical intercalation,} perturbations to the electronic structure of metallic systems are confined within few atomic layers even at the largest applied electric fields \cite{PiattiApsusc2018NbN}. Hence, proximity effect between the perturbed surface layer and the bulk strongly hampers any $T_c$ modulation \cite{PiattiPRB2017, UmmarinoPRB2017}. Thus, either further experiments on thinner films (1-2 unit cells at most), or a full theoretical treatment of proximity effect in ion-gated BaFe\ped{2}(As,P)\ped{2} thin films, will be required to elucidate the issue in this class of compounds.

\section{Conclusions}

In summary, we performed ionic gating measurements on ultrathin ($10$ nm) films of optimally-doped BaFe\ped{2}(As,P)\ped{2} grown on MgO substrates via molecular beam epitaxy. We controlled the {\color{blue}charge doping} at the film surface by employing an optimized polymer electrolyte designed to {\color{blue}reduce} undesirable electrochemical interactions with the sample. The resulting modulations to the resistivity were found to be compatible with a {\color{blue}tuning of the charge doping with a dominant electrostatic contribution, and }with a scaling on the induced charge density consistent with an asymmetric efficiency as surface scattering centers between cations and anions.
At low temperatures, the SC transition temperature was suppressed both upon electron and hole doping, indicating that SC is fully optimized by P substitution and any further deviation from this optimal condition via {\color{blue}ionic gating} is detrimental to the SC state. Additionally, we showed that {\color{blue} the gate-induced charge doping} leads to a broadening of the resistive transition. This indicates that, unlike in the case of thin films of standard BCS superconductors, gate-induced modulations to the SC order parameter in Ba-122 may not be uniform across the entire film thickness. 
{\color{blue}Our results provide valuable insights in the optimization of the SC transition temperature in the 122 family of iron-based superconductors by means of charge doping, laying a foundation for more advanced studies. Among these, we consider especially interesting the assessment of the effects of the gate-induced charge doping on the SC properties of underdoped films and the investigation of contributions to the gating mechanism in this class of materials beyond pure charge doping, such as field-induced distortions of the crystal lattice and protonation; these could be achieved by a combination of \textit{ab initio} calculations and direct probing of the crystal structure via X-ray diffraction measurements performed \textit{in operando}.}

\section*{Acknowledgments}
This work was partially supported by the Program for Advancing Strategic International Networks to Accelerate the Circulation of Talented Researchers (no. R2605) from the Japan Society for the Promotion of Science.

\end{document}